\documentclass[sw]{iosart2x}

\usepackage{svg}
\usepackage{multirow}
\usepackage{float}
\usepackage{amsmath}
\usepackage[strings]{underscore}

\pubyear{0000}
\volume{0}
\firstpage{1}
\lastpage{1}

\begin{document}

\begin{frontmatter}

\title{A dual approach to ShEx visualization with complexity management}
\runtitle{A dual approach to ShEx visualization with complexity management}

\begin{aug}
\author[A]{\inits{J.}\fnms{Jorge} \snm{Alvarez-Fidalgo}\ead[label=e1]{UO258524@uniovi.es}
\thanks{Corresponding author. \printead{e1}.}}
\author[A]{\inits{J.E.}\fnms{Jose Emilio} \snm{Labra-Gayo}\ead[label=e2]{labra@uniovi.es}}
\address[A]{Dept. of Computer Science, \orgname{University of Oviedo}, \cny{Spain}\printead[presep={\\}]{e1,e2}}
\end{aug}


\begin{abstract}
Shape Expressions (ShEx) are used in various fields of knowledge to define RDF graph structures. ShEx visualizations enable all kinds of users to better comprehend the underlying schemas and perceive its properties. Nevertheless, the only antecedent (RDFShape) suffers from limited scalability which impairs comprehension in large cases. In this work, a visual notation for ShEx is defined which is built upon operationalized principles for cognitively efficient design. Furthermore, two approaches to said notation with complexity management mechanisms are implemented: a 2D diagram (Shumlex) and a 3D Graph (3DShEx). A comparative user evaluation between both approaches and RDFShape was performed. Results show that Shumlex users were significantly faster than 3DShEx users in large schemas. Even though no significant differences were observed for success rates and precision, only Shumlex achieved a perfect score in both. Moreover, while users' ratings were mostly positive for all tools, their feedback was mostly favourable towards Shumlex. By contrast, RDFShape and 3DShEx's scalability is widely criticised. Given those results, it is concluded that Shumlex may have potential as a cognitively efficient visualization of ShEx. In contrast, the more intricate interaction with a 3D environment appears to hinder 3DShEx users.
\end{abstract}


\end{frontmatter}

\section{Introduction}\label{s1}
Shape Expressions (ShEx) \cite{shex} was proposed in 2014 as a language for RDF\footnote{http://www.w3.org/RDF/} data validation. By allowing to define RDF graph structures, it enables data producers and consumers to settle in a common ground and avoid inconsistencies. Since RDF brings together users from various branches of human knowledge, ShEx is employed in a variety of different contexts. E.g., ShEx is used to validate the RDF representation of FHIR\footnote{https://www.hl7.org/fhir/rdf.html}, a standard for health care data exchange.

This implies that users do not necessarily have to be familiar with textual programming languages, resulting in a steep learning curve. One possible solution to such problem is the use of \textbf{visualizations}. They enable users to comprehend sheer amounts of data in a efficient manner and allows for better perception of emergent properties, errors and patterns \cite{ivpfd}. 

The only precedent as far as ShEx is concerned is RDFShape \cite{rdfshape}, being capable of generating UML-like\footnote{https://www.omg.org/spec/UML/} class diagrams for a subset of the language. Alas, it suffers from limited scalability as well as a degree of symbol overload which may affect its semantic transparency. Therefore, the information conveyed may be cognitively inefficient, particularly in larger use cases.

Other visualisations in the Semantic Web ecosystem formulate different solutions to the problem of developing a comprehensible visual notation, with varying degrees of success. However, the aforementioned issue of scalability -also referred to as complexity management- is rarely addressed. At most, a few -such as WebVOWL\footnote{https://service.tib.eu/webvowl/}- provide automatic mechanisms for reducing the number of elements displayed, but no choice is given to the user about the specifics.

Thus, the main \textbf{contribution} of this work lies in the proposal of a visual notation for ShEx which aims to be cognitively efficient -with an emphasis on complexity management-, analysing the perceptual implications of its materialisation in both a 2D plane and 3D space.

The rest of the paper is structured as follows. A motivating example is provided in Section \ref{s2}. In Section \ref{s3}, background information about cognitive implications of visual design is provided, as well as the state of the art about visualization tools in the Semantic Web. The proposed visual notation and both approaches to it are exposed in Section \ref{s4}. The implementation of the corresponding prototypes is discussed in Section \ref{s5}. User evaluation methodology, results and discussion are provided in Section \ref{s6}. Finally, conclusions and future work are discussed in Section \ref{s7}.

\section{Motivating example}\label{s2}
The Wikidata GeneWiki \cite{genewiki} project aims to use Wikidata as a semantic framework to manage and disseminate biomedical data. To that end, it describes a knowledge graph about such entities and their relationships. Our ShEx motivating example\footnote{https://github.com/weso/sparkwdsub/blob/master/examples/genewiki.shex} defines this graph structure. Given its abundant number of elements, it poses a challenge for proper visualization.

\begin{figure}[h]
\includegraphics[scale=0.25]{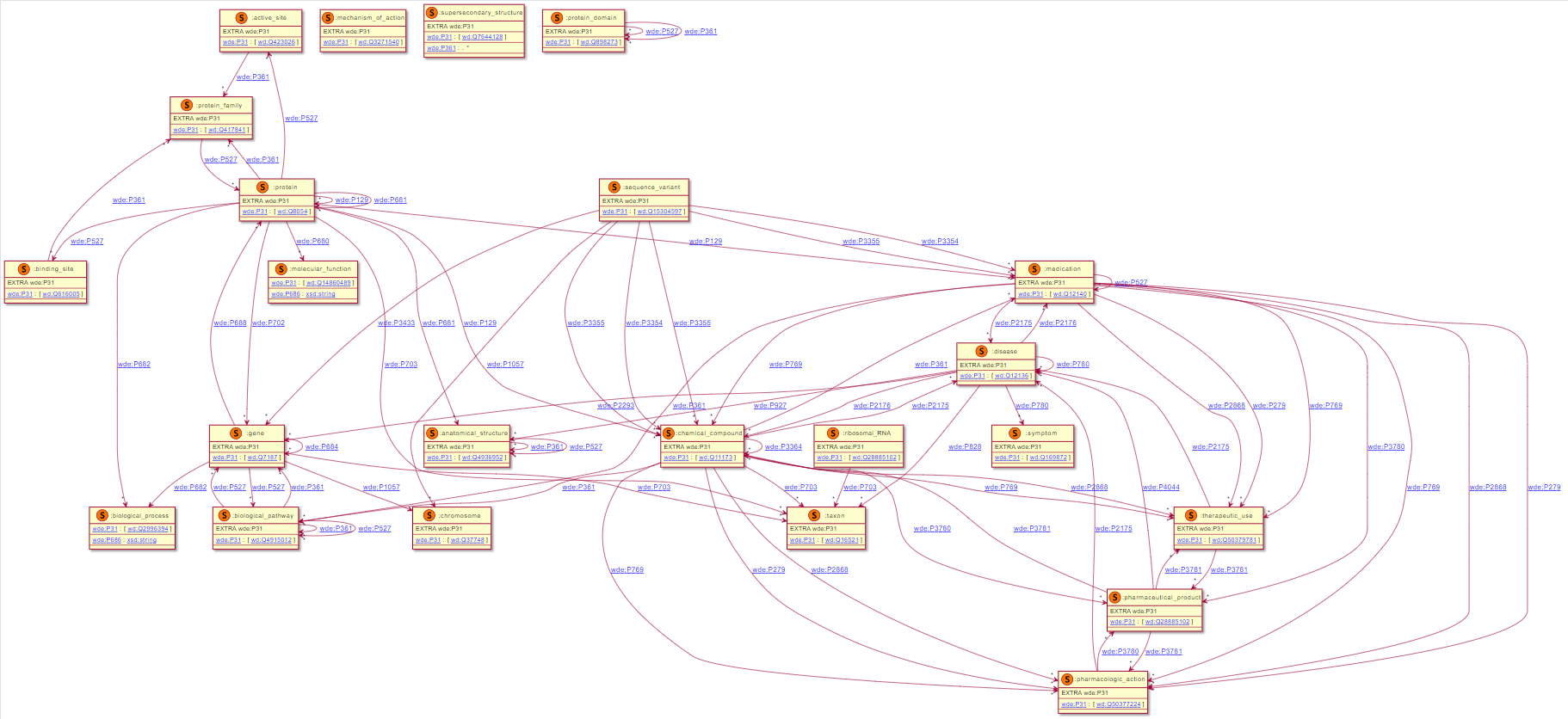}
\caption{Genewiki ShEx visualization in RDFShape.}\label{f1}
\end{figure}

RDFShape's visual representation (DOT)\footnote{https://rdfshape.weso.es/link/16520062624} generates a class diagram with 23 classes and over 70 relationships (see Fig. \ref{f1}). Besides the cognitive implications of processing a large network -which will be discussed later-, common scalability issues may be observed. Key sections of the diagram become filled with relationships and difficult to discern between each other. Therefore, it is a suitable testing ground for testing complexity management mechanisms and the cognitive efficiency of the visual notation.

\section{State of the Art}\label{s3}
In this section, (i) cognitive implications of visual notation design and (ii) visualizations in the Semantic Web are discussed.

\subsection{Cognitive implications of visual notation design}\label{s3.1}
\textbf{Cognitive load theory} "is concerned with the manner in which cognitive resources are focused and used during learning and problem solving" \cite{chandlersweller}. It describes the impairment of understanding that takes place when learning procedures lead to further cognitive processes. A distinction is made between intrinsic and extraneous cognitive load; the former is due to the inherent complexity of the information, while the latter is generated because of the manner in which such information is presented. Those phenomena are related in such a way that the consequences of extraneous cognitive load may only be noticeable when dealing as well with intrinsic cognitive load caused by high element interactivity \cite{sweller94}.

D. Moody describes in his \textbf{Physics of Notations} (PoN) theory a series of principles for designing cognitively effective visual notations \cite{moody}: \emph{semiotic clarity, perceptual discriminability, semantic transparency, complexity management} (reduction of extraneous cognitive load), \emph{cognitive integration, visual expressiveness, dual coding, graphic economy and cognitive fit}. In the last decade, PoN has become a widely used standard for notation design to the detriment of competing approaches \cite{reviewpon}. 

Various criticisms have been stated about PoN. The operationalization of said principles ranges from objective measures -semiotic clarity is a 1:1 correspondence- to subjective evaluations -the "suggestion of meaning" implied by semantic transparency may only be determined by empirical means- \cite{operation}. This implies a degree of user involvement usually lacking in its application \cite{ponuser}. Subsequent proposals were made in order to improve such operationalization, either partially \cite{oppar} or completely \cite{ponfram}.

On a different note, the impact of 3D visualizations in cognitive load may be closely related to spatial ability \cite{learning3d}. I.e. subjects with high spatial ability perceive their cognitive load as low and vice versa. Further research shows that this effect is exacerbated when dealing with static visualizations; dynamic interactions providing a compensating effect for low spatial ability learners \cite{spatial}. 

\subsection{Visualizations in the Semantic Web}\label{s3.2}
As stated in the introduction, \textbf{RDFShape} provides the only visualization currently available for Shape Expressions\footnote{https://rdfshape.weso.es/shexInfo}. It generates a bidimensional graph in which UML-like boxes symbolize shapes and directional arrows represent references to other shapes. No interactive actions nor complexity management mechanisms are provided.

Further work has been carried out for the closely related Shapes Contraints Language (SHACL)\footnote{https://www.w3.org/TR/shacl/} in the form of visual editors. Arndt et al. implemented an Ontopad-based\footnote{https://github.com/AKSW/OntoPad} tool which allows for composing a SHACL visual data model \cite{ontopad}. Most of the interaction is done through a textual interface; new elements must be dragged into a canvas to become part of the visualization. Users may perform a few tasks on the visualization, such as linking properties.  

Lieber et al. define both UML-based and VOWL-based visual notations to represent RDF constraints and implement them in \textbf{UnSHACLed},a SHACL visual editor \cite{unshacled}. Empirical tests showed no significant difference in error rates between the approaches. Nevertheless, the majority of users did prefer the VOWL-based notation. The authors acknowledge the need for complexity management mechanisms, but it is considered out of their scope.

\textbf{VOWL} \cite{vowl} is a visual notation for representing OWL\footnote{https://www.w3.org/TR/owl-features/} ontologies, with two implementations available: WebVOWL and ProtégéVOWL. WebVOWL makes use of a force graph, allowing for user interaction with the positioning of elements. Moreover, it provides a complexity management tool: a collapsing feature which reduces the number of elements on screen, even though it leaves no choice to the user over the specifics.  Nonetheless, this does not prevent the overlapping of a large number of relationships between two nodes.

This overlapping issue has been a recurring problem in the history of visualizations in the semantic web. As far back as the year 2000, a possible solution emerged: to represent semantic graphs in three dimensions. With such purpose tools as \textbf{UNIVIT} \cite{univit} and \textbf{NV3D} were implemented; alas, their visual notations were too dependant on the arbitrary combination of visual variables (shape, color...) to efficiently convey complex data \cite{vissem2}.

A decade later, \textbf{X3D-UML} was proposed as a 3D UML implementation, particularly focused on state machine diagrams \cite{x3d}. It consists of a number of interconnected planes in a tridimensional space, each one displaying a 2D UML diagram. Thus, it is rather an intermediate solution.

\section{Proposal}\label{s4}
This proposal consists of a UML class diagram-like visual notation in order to graphically represent ShEx, with two alternative approaches: a 2D diagram (\textbf{Shumlex}) and a 3D directed graph (\textbf{3DShEx}).

\subsection{Visual notation}\label{s4.1}
The proposed visual notation is displayed in Table \ref{t1}. Its design rationale is structured according to PoN's principles, exposed hereunder.
\subsubsection{Semiotic Clarity}\label{semioticclarity}
Semiotic clarity is sacrificed in favour of both semantic transparency and graphic economy. Firstly, the notation incurs in a deliberate case of \textbf{symbol deficit}, since there exist a number of semantic constructs without a unique visual construct mapped to it. Node constraints are displayed textually inside shapes, much like attributes in UML classes. By doing so, it is expected to take advantage of the widely recognized UML class diagram notation to convey information to a broader audience.

Secondly, some semantic constructs employ the same visual construct (\textbf{symbol overload}) with textual differentiation. Since those semantic constructs are conceptually similar (e.g. conjunction and disjunction) it is hoped to achieve graphic economy without disrupting clarity.

\subsubsection{Perceptual discriminability}
In order to objectively ascertain the ease of discrimination between symbols, both a metric and a threshold of dissimilarity between two graphical symbols have to be defined \cite{ponfram}. To that end, the proposal from \cite{oppar} is modified slightly in order to calculate the average of the following criteria: visual distance (VD), redundant coding (RC), perceptual pop-out (PPO) and textual differentiation (TD). These are normalized to an interval of [0, 1], in such a way that 0 denotes null discriminability and 1 compliance with all criteria. 0.5 is chosen as the threshold of dissimilarity.

For brevity's sake, the details of the modifications and calculations are exposed in Appendix B. The values obtained were VD = 0.47, RC = 0.29, PPO = 1 and TD = 0.5. This results in an average value of 0.57, therefore demonstrating the positive perceptual discriminability.

\subsubsection{Semantic transparency}
By shaping the notation to resemble a UML class diagram, the objective is to increase its semantic transparency, particularly to novice users. At least, semantic translucency is expected, so that the visual constructs provide a cue to its meaning by association. Nonetheless, this cannot be ascertained until a user evaluation is performed, given its aforementioned subjectivity.

\subsubsection{Complexity management}
Complexity management is approached in different ways. Shumlex takes inspiration from the modularization utility implemented by GraphQL Voyager\footnote{https://apis.guru/graphql-voyager/}, which by clicking on one of the classes highlights only that class and its relations with its neighbours, drastically reducing the others' visibility. A modification of this concept is proposed such that it is cumulative; that is, clicking on a second class does not change the focus to that class, but adds it to the highlighted set. 

\begin{table}[H]
\caption{Proposed visual notation.} \label{t1}
\begin{tabular}{|l|l|c|l|}
\hline
\textbf{Feature} & \textbf{\{PLACEHOLDER\}}                                                                                                                                  & \textbf{Visual representation} & \textbf{Example}                                                                                                                                                               \\ \hline
TripleConstraint & \begin{tabular}[c]{@{}l@{}}\textless{}Property\textgreater \\ \textless{}NodeConstraint\textgreater \\ \textless{}Cardinality\textgreater{};\end{tabular} & \multirow{5}{*}{\centering
\def\svgscale{0.6}
\raisebox{-2.4cm}{\includegraphics[scale=0.3]{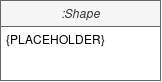}}  }              & \begin{tabular}[c]{@{}l@{}}:User \{ \\      :name xsd:string ?; \\  \}\end{tabular}                                                                                            \\ \cline{1-2} \cline{4-4} 
EachOf           & \textless{}TripleConstraint\textgreater{}+                                                                                                                &                                & \begin{tabular}[c]{@{}l@{}}:User \{ \\      schema:name xsd:string; \\      schema:age xsd:string;\\  \}\end{tabular}                                                          \\ \cline{1-2} \cline{4-4} 
\begin{tabular}[c]{@{}l@{}}Top-level \\ Nodekind\end{tabular}    & nodeKind: \textless{}NodeKind\textgreater{}                                                                                                               &                                & :HomePage IRI                                                                                                                                                                  \\ \cline{1-2} \cline{4-4} 
Extra            &                EXTRA <ValueSet>                                                                                                                                           &                                & \begin{tabular}[c]{@{}l@{}}:User EXTRA schema:name \{            \\     schema:name xsd:string ; \\  \}\end{tabular}                                                           \\ \cline{1-2} \cline{4-4} 
Closure            &                CLOSED                                                                                                                                           &                                & \begin{tabular}[c]{@{}l@{}}:User CLOSED \{\} \end{tabular}                                                           \\ \hline
ShapeRef         &                                                                                                                                                           &         
\centering
\def\svgscale{0.6}
\raisebox{-0.3cm}{\includegraphics[scale=0.3]{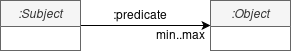}}                           & \begin{tabular}[c]{@{}l@{}}:User \{ \\      schema:worksFor @:Company ;  \\  \}\end{tabular}                                                                                   \\ \hline
ShapeAnd         & AND                                                                                                                                                       & \multirow{3}{*}{\centering
\def\svgscale{0.6}
\raisebox{-3.3cm}{\includegraphics[scale=0.3]{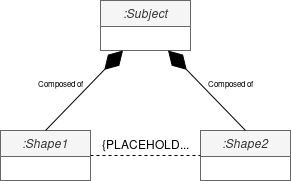}}}              & \begin{tabular}[c]{@{}l@{}}:User \{  \\     schema:name xsd:string ;  \\ \}  AND \{  \\     schema:age xsd:integer  \\  \}\end{tabular}               \\ \cline{1-2} \cline{4-4} 
ShapeOr          & OR                                                                                                                                                        &                                & \begin{tabular}[c]{@{}l@{}}:User \{  \\     schema:name xsd:string  \\  \} OR \{  \\     schema:givenName xsd:string ; \\      \}\end{tabular} \\ \cline{1-2} \cline{4-4} 
OneOf            & OneOf                                                                                                                                                     &                                & \begin{tabular}[c]{@{}l@{}}:User \{ \\     :name xsd:string; | \\  ( \\     :givenName xsd:string +; \\     :familyName xsd:string; );  \}\end{tabular}                     \\ \hline
ShapeNot         & NOT                                                                                                                                                          & \multirow{2}{*}{\centering
\def\svgscale{0.6}
\raisebox{-1.7cm}{\includegraphics[scale=0.3]{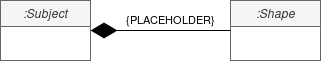}}}              & \begin{tabular}[c]{@{}l@{}}:NoName Not \{    \\     schema:name .  \\  \}\end{tabular}                                                                                         \\ \cline{1-2} \cline{4-4} 
Labelled         & Composed of                                                                                                                                                         &                                & \begin{tabular}[c]{@{}l@{}}:User \{ \\     \$:name ( \\        :name .; \\        :email IRI; \\     ) \\  \}\end{tabular}                                                     \\ \hline
\end{tabular}
\end{table}

As far as 3DShEx is concerned, an \emph{intelligent zoom} \cite{intzoom} will be implemented. This implies additional interactivity besides the common zooming. By default, nodes will display only the shape identifier. When interacted with -opened-, nodes will expand to reveal the pertinent restrictions (\emph{black boxing}). Further interaction -closing- will revert it to its initial state. As an additional complexity management tool, a simple collapsing function is proposed which on demand shows only the desired node and its neighbours. Both provide a layer of abstraction.

\subsubsection{Cognitive integration}
Given the fact that no multiple diagrams are used to represent a dataset, this principle does not apply.

\subsubsection{Visual expresiveness}
PoN builds upon Bertin's list of visual variables \cite{bertin}: shape, texture, brightness, size, color, orientation and planar variables. The proposed notation makes use of the following: shape, texture and brightness. Hence, it lies in a middle ground between visual one-dimensionality and visual saturation. Moody claims that most diagrams in software engineering are visually one-dimensional \cite{moody} and therefore a higher degree of discriminability is achieved, as discussed earlier.

\subsubsection{Dual coding}
Every visual representation is complemented by text which provides a cue to its meaning.

\subsubsection{Graphic economy}
As previously stated, symbol deficit was introduced in order to reduce graphic complexity. Consequently, the number of graphical symbols is 4. Such is an inferior value to the 7±2 processing capacity limit proposed by Miller \cite{miller}, suggested in \cite{ponfram} as a reference for this principle.

\subsubsection{Cognitive fit}
The aim of the aforementioned binary approach is to maximize cognitive fit, each prioritizing different necessities. Shumlex aims to be closer to the UML class diagram spec, with the intention of being accessible to a wider audience not necessarily familiar with the technical details. Thus, all information is initially available as it would be in a common diagram, making for a more constraint-focused visualization in contrast to a relationship-focused 3DShex.

On the contrary, 3DShex is of a more experimental nature, which by interactively presenting the same information in a tridimensional space aims to further analyse the cognitive implications in its audience and the potential benefits it may bring. The details of shapes are concealed behind a layer of abstraction, thus giving greater importance to the diagram as a whole.

\section{Implementation}\label{s5}
In this section, the elaboration of the prototypes for both approaches to the visual notation is described.
\subsection{Shumlex}\label{s5.1}
In order to build the visualization Mermaid\footnote{https://mermaid-js.github.io/mermaid/}, a Javascript library for text-based generation of various diagrams, is used. Therefore, the architecture of the prototype is as follows:
\begin{enumerate}
  \item A \textbf{conversion} engine which receives a ShEx input and generates the Markdown-like syntax that Mermaid requires. Given the fact that Mermaid does not accept a variety of symbols used in ShEx, it is necessary to use alternatives. For instance, the use of colons is not allowed; the prefixed term ":User" would have to be codified as "\_User".
  \item A \textbf{visualization} module which invokes the library with the previous outcome in order to generate a SVG. Once displayed, the sanitized texts are substituted by the original ones.
  \item A \textbf{post-processing} module which implements the complexity management funcionality. It assigns to every class in the diagram an event which, on click, lowers to a minimum the opacity of every element except that very class, its relationships and the targets of these. There exist a couple of exceptions: a) it won't obscure the already highlighted elements and b) it will reverse the effect if it has already been applied to said class. Furthermore, hovering any label will check the existence of the entity in Wikidata and display its meaning as a tooltip. The purpose of this is to increase comprehension of commonly used, Wikidata related Shape Expressions, in which shapes and predicates are semantically opaque (e.g. wd:Q42944 refers to CERN).
\end{enumerate}

\begin{figure}[h]
\includegraphics[scale=0.3]{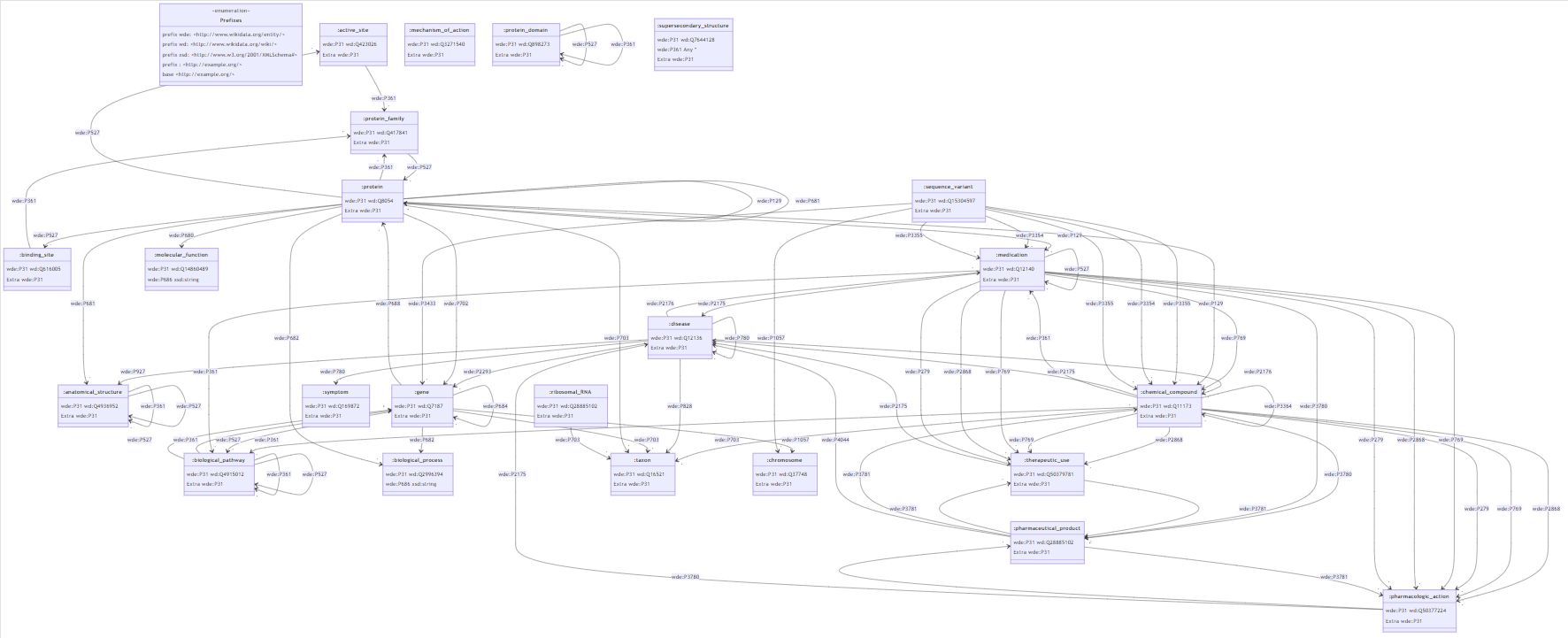}
\caption{Genewiki ShEx visualization in Shumlex.}\label{f2}
\end{figure}

\begin{figure}[h]
\includegraphics[scale=0.3]{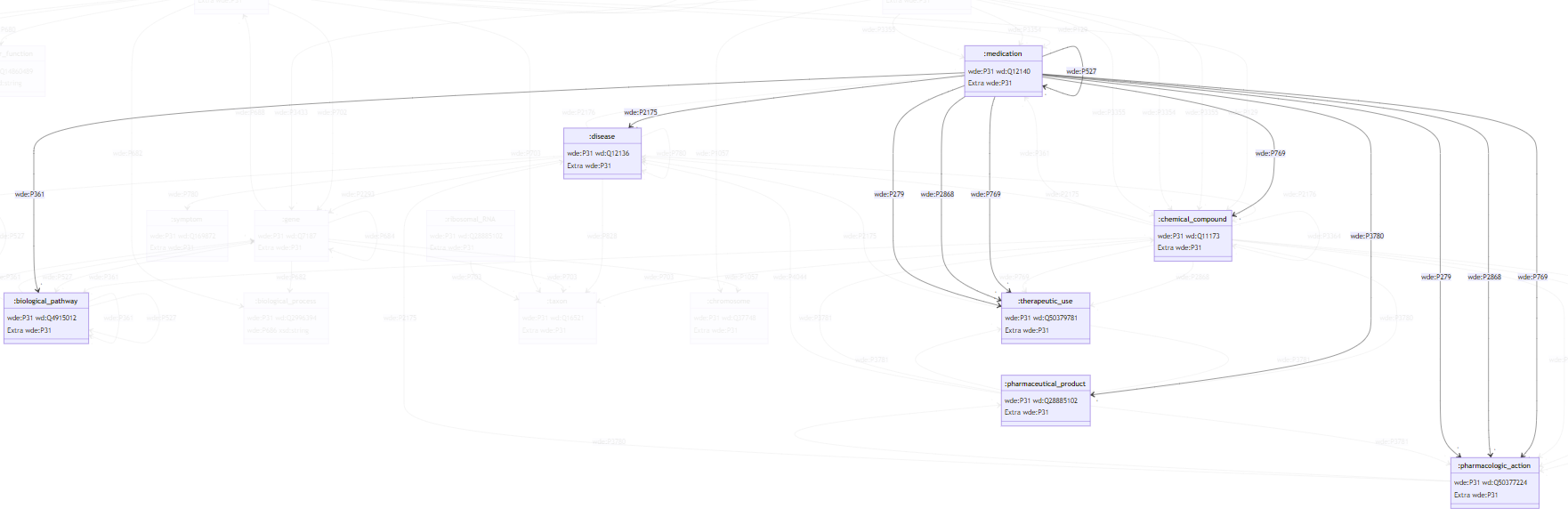}
\caption{Genewiki ShEx visualization with reduced complexity in Shumlex.}\label{f3}
\end{figure}

The application of Shumlex to the motivating example is shown in Fig. \ref{f2}. Despite the fact that relationships are more spaced out than in RDFShape's visualization, areas with high concentrations of elements remain cognitively overloaded. As shown in Fig. \ref{f3}, the complexity management mechanism allows for a limited display of the desired components. In the provided example, focus is on \emph{:medication}, thus highlighting its relationships with other shapes.

This prototype is freely available at http://www.weso.es/shumlex/.

\subsection{3DShex}\label{s5.2}
For the implementation of this prototype 3D Force Graph\footnote{https://github.com/vasturiano/3d-force-graph} (3DFG), a NodeJS library to represent graph data in a tridimensional space, is used. The architecture of the prototype is as follows:
\begin{enumerate}
  \item A \textbf{conversion} engine which receives a ShEx input and generates the JSON data that 3DFG requires. Besides the required parameters, additional information is included to facilitate the next phase. 
  \begin{enumerate}
  \item \emph{List of constraints of a node}. Information to be displayed on demand, equivalent to the class attributes in UML.
  \item \emph{Name and cardinality of a relationship}. 
  \item \emph{Curvature of a link}. Curved links allow for distinct relationships between a pair of nodes, while straight links only enable one to be displayed clearly. Therefore, references between shapes (:User :works @:Company) require the former since there may be any number of them. On the contrary, compositional relationships (ShapeAnd, ShapeOr, OneOf, ShapeNot and Labelled in Table \ref{t1}) are unique to the source shape and thus are able to be represented by straight links.
  \item \emph{Arrow head}. As displayed in the notation, there are three posibilities: none, arrow or diamond.
  \item \emph{Rotation}. As previously mentioned, curved links allow for an arrangement free of the overlapping described in Section \ref{s3.2}. However, by default, links are displayed in the same position. 3DFG allows for a rotation value -taking the node as the centre of a circunference- to be specified, but the calculations are left to the user.
  Hence, every link occurrence for each node pair is registered and the circunference is divided in equal parts. Moreover, the source of the link should be taken into account, since from the perspective of each circunference the angle will be different for a certain position (e.g. $\pi$ in the source node equals to $2\pi$ in the target).
\end{enumerate}
  \item A \textbf{visualization} module which makes use of the previous information to invoke the library. HTML objects are utilized to build the nodes, so their contents can be customized as well as dynamic behaviour assigned (constraints are hidden by default). The following functionalities are enabled:
  \begin{enumerate}
  \item \emph{Highlight on hover}. Both links and nodes possess this property; in the case of the latter, its neighbours and the corresponding relationships are emphasized as well. In links, moving particles are shown to reinforce the direction.
  \item \emph{Details}. When clicking a node, all its constraints are displayed in a expanded box. Another click reverts it to its original state.
  \item \emph{Collapsing}. Right clicking a node displays a reduced graph, composed of such node and its neighbours.
  \item \emph{Wikidata tooltips}. As in Shumlex.
\end{enumerate}
\end{enumerate}

\begin{figure}[h]
\includegraphics[scale=0.35]{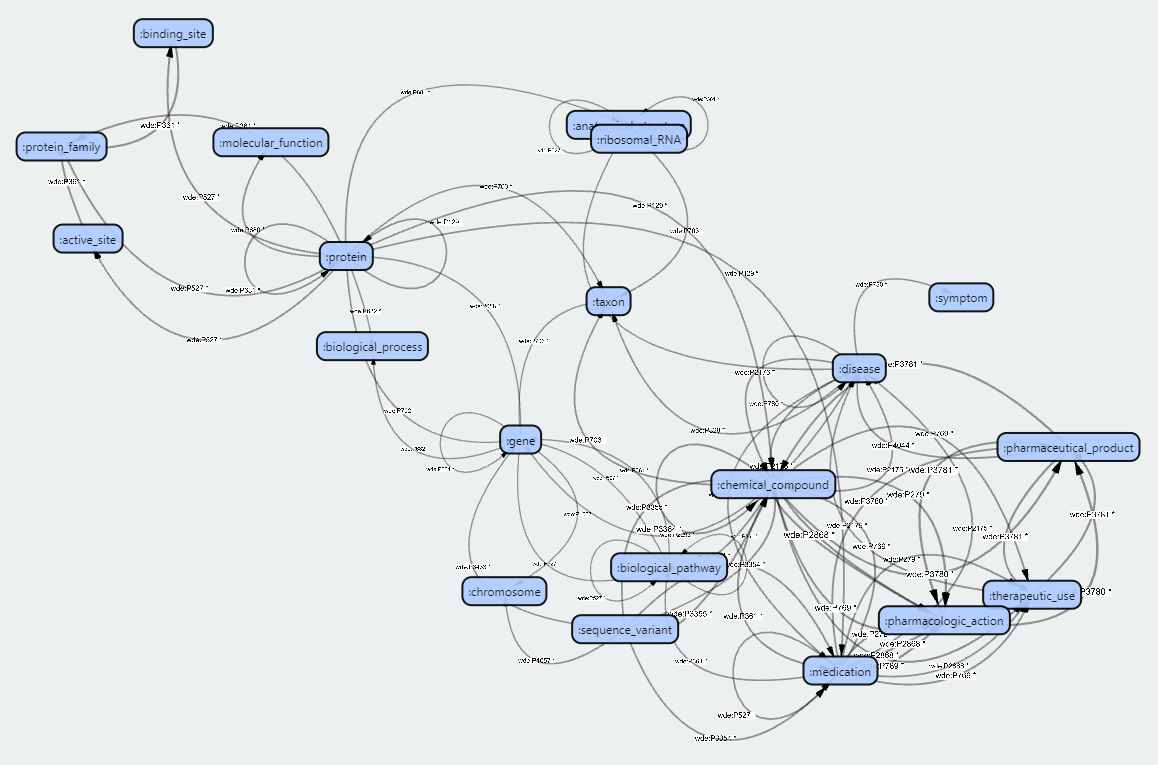}
\caption{Genewiki ShEx visualization in 3DShEx.}\label{f4}
\end{figure}

\begin{figure}[h]
\includegraphics[scale=0.35]{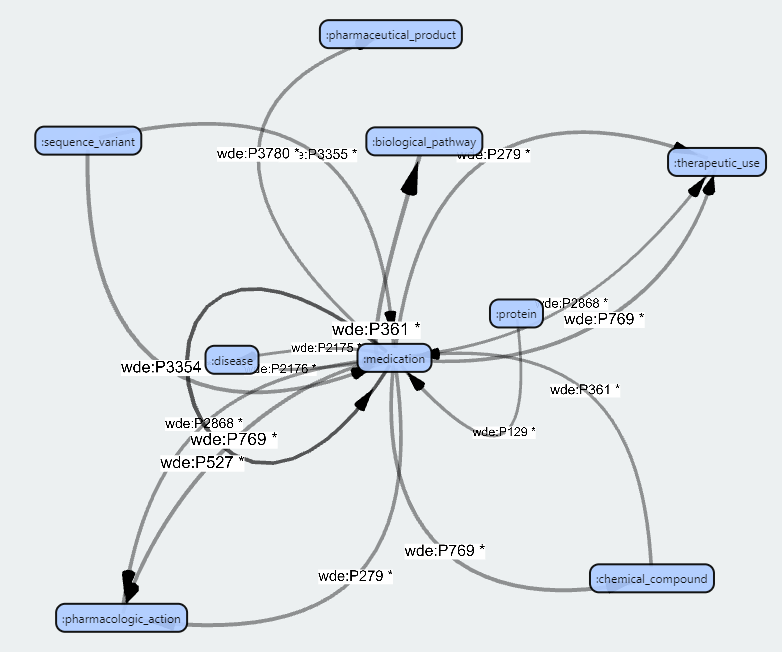}
\caption{Genewiki ShEx visualization with reduced complexity in 3DShEx.}\label{f5}
\end{figure}

The application of 3DShEx to the motivating example is shown in Fig. \ref{f4}. Even though static images do little for its comprehension -since it may be examined from any position- it is clear that clusters of highly interdependent shapes excess working memory limits. The complexity management mechanism is thus applied to \emph{:medication} yet again. As shown in Fig. \ref{f5}, a much smaller graph is displayed, composed by the desired element and its neighbours.

This prototype is freely available at http://www.weso.es/3dshex/.

\section{Evaluation}\label{s6}
In order to test the proposed notation, an experiment was carried out based in the one conducted in \cite{shexml}. The methodology employed, the results obtained and their discussion are detailed in the following subsections. Datasets, questionnaires, manuals and anonymized results are freely available at https://github.com/fidalgoLXXVI/shex-visualization-paper.
\subsection{Methodology}\label{s6.1}
 This user study follows a between-subjects design, in which each participant is exposed to a single tool and asked to perform a few measured tasks. Both a quantitative and a qualitative analysis are conducted. Hereunder, the methodology of this experiment is discussed in greater detail.
\subsubsection{Procedure}
The experiment is divided in the following steps:
\begin{enumerate}
    \item \textbf{Preliminary questionnaire}. Subjects are inquired about background and self-assessment of relevant skills -such as knowledge in UML or spatial ability-.
    \item \textbf{Tool description}. A brief manual is provided to participants which briefly describes the operation and features of the corresponding tool. The selected tools for the experiment were \emph{RDFShape, Shumlex} and \emph{3DShEx}. Thus, the different approaches to the notation may be compared to each other as well as to the existing solution.
    \item \textbf{Main questionnaire}. A series of tasks on the test cases are requested to the participants. By means of the mandated tool, subjects must try to perform those while their interactions are measured by a timer. Each test case comprises the following tasks, which aim to ascertain the user's ability to navigate the diagram and comprehend the various semantic equivalences.
    \begin{enumerate}
    \item \emph{Find a shape by name}.
    \item \emph{Find a shape with a specific constraint}. 
    \item \emph{List within-node constraints of a shape}.
    \item \emph{Find a reference between two shapes}.
    \item \emph{Determine subject and object of a reference.}. 
    \item \emph{List all neighbours of a shape}. 
\end{enumerate}
    \item \textbf{Follow-up questionnaire}. A number of questions based in the Likert scale are asked to the participants in order to perform the qualitative analysis. Those allow us to obtain a number of variables: \emph{general satisfaction level, ease of use, learnability, semantic transparency, applicability, error proneness, scalability, complexity management, understanding of constraints and  understanding of references}.
\end{enumerate}

\subsubsection{Sample}
The sample consisted of 13 students of the MSc in Web Engineering at University of Oviedo. This experiment took place in the last day of a course in semantic web, where they were taught the basics of technologies such as RDF or ShEx. Most participants share a similar demographic as well as academic background, with a bachelor's degree in Computer Science. According to self-assessment results a) 92.3\% have either medium or high knowledge of UML, b) 84.6\% have basic knowledge of RDF, c) 69.2\% have basic knowledge of ShEx and 15.4\% declare no knowledge on the subject and d) 69\% have high spatial ability while the rest declare medium spatial ability.

\subsubsection{Test cases}
Two test cases are used in this experiment. The first one is based on the WebIndex ShEx schema proposed in \cite{valrdf}, "one of the earliest practical applications of ShEx". Modifications have been made in order to reflect all the features reflected in the visual notation, thus including logical operations and composition. The \emph{OneOf} constraint is removed since RDFShape's current version hasn't implemented it yet. This schema features a few shape references, with greater focus on other node constraints.

The second one is the Genewiki schema, as featured in Section \ref{s2}. It has approximately three times the shapes of the former and over 70 shape references, while other constraints are a scarce ocurrence.

Hence, according to the cognitive implications layed out in Section \ref{s3.1}, participants would be confronted with \textbf{distinct cognitive loads}. The first test case has a higher intrinsic load, given the greater inherent complexity of using complex semantic features such as conjunctions and composition while having few elements. On the contrary, the second test case has little implicit complexity -most are simple references to other shapes- but its large quantity of elements causes diagrammatic complexity upon display.

\subsubsection{Threats to validity}
Taking as reference the list of threats to both internal and external validity proposed in \cite{creswell}, the following have been identified:
\paragraph{Selection.} Participants may share certain characteristics which predispose them towards the same results, especially given the common background. In order to address this, subjects are distributed randomly among the experimental groups so that those characteristics may be equally distributed.
\paragraph{Testing.} Participants may become familiar with the test cases and remember responses for later tasks. In order to mitigate this, special care is taken to use different fragments of the schema and avoid repetitions.
\paragraph{Interaction of selection and treatment.} Because of the limited variety of the participants, generalization to individuals of other contexts may not be possible. Hence, claims about the universality of the results must be restricted. However, given the highly specialized nature of the contribution, this issue is lessened.

\subsubsection{Analysis}
Both quantitative and qualitative results were collected and anonymised. From those, the following variables are calculated for each test case: \textbf{elapsed time}, \textbf{success rate} and \textbf{precision}. Elapsed time ($T_c$) is the total time spent for a given test case. Success rate ($S_c$) is calculated as the number of correct answers divided by the number of questions. Precision ($P_c$) is calculated as the division of minimum elapsed time of all participants by current student's elapsed time, multiplied by the success rate. This measure gives an insight on the swiftness of participants while taking into account their effectiveness. Hence, given a test case \emph{c} and a student \emph{sn}:
\begin{equation*}
P_{csn} := \frac{min(\{T_{cs1},...,T_{csn}\})}{T_{csn}} \cdot S_{csn}
\end{equation*}
R 4.2.0 is used for the statistical analysis. Comparisons between the three groups are made by means of a One-Way ANOVA whenever assumptions are met, removing outliers if necessary. Otherwise, Kruskal-Wallis is used.

\subsection{Results}\label{s6.2}
Descriptive statistics of the quantitative results for the first test case are shown in Table \ref{t2}. Shumlex mean scores are consistently better than RDFShape's, and those better than 3DShEx's. Nonetheless, those differences between the three groups are not statistically significant for any of the variables: \emph{F(2,8)=1.1; p=0.377},  \emph{F(2,10)=1.67; p=0.236} and \emph{F(2,9)=1.29; p = 0.32} for T, S and P respectively.

\begin{table}[H]
\caption{Descriptive statistics for test case 1 results.} \label{t2}
\begin{tabular}{llllll}
\hline
\textbf{Measure}                                      & \textbf{Group} & $\mathbf{\overline{x}}$ & \textbf{s} & \textbf{max} & \textbf{min} \\ \hline
Elapsed seconds                   & 3DShEx                          & 256.2                   & 66.55                       & 355                           & 210                           \\
                                  & RDFShape                        & 210.2                   & 119.62                      & 411                           & 95                            \\
                                  & Shumlex                         & 196                     & 95.63                       & 302                           & 73                            \\
Success rate                      & 3DShEx                          & 0.667                   & 0.136                       & 0.833                         & 0.5                           \\
                                  & RDFShape                        & 0.7                     & 0.139                       & 0.833                         & 0.5                           \\
                                  & Shumlex                         & 0.833                   & 0.136                       & 1                             & 0.667                         \\
Precision                         & 3DShEx                          & 0.204                   & 0.077                       & 0.29                          & 0.103                         \\
                                  & RDFShape                        & 0.311                   & 0.197                       & 0.64                          & 0.118                         \\
                                  & Shumlex                         & 0.441                   & 0.380                       & 1                             & 0.161                         \\ \hline
\end{tabular}
\end{table}

Descriptive statistics of the quantitative results for the second test case are shown in Table \ref{t3}. Mean score comparisons show the same relationship between groups as before. However, in this case there are significant differences between the three groups in elapsed times (\emph{H(2)=6.05; p=0.048; $\eta^2$=0.405}). Dunn post-hoc determined significant differences in elapsed times between Shumlex and 3DShEx (\emph{p=0.014}). While 75\% of Shumlex users achieve lower times than every RDFShape user, overall differences are not significant (\emph{p=0.242}).

As success rate and precision are concerned, there are no significant differences between groups in the second test case (\emph{H(2)=1.78; p = 0.41 and F(2,10)=2.43; p=0.137}).

\begin{table}[H]
\caption{Descriptive statistics for test case 2 results.} \label{t3}
\begin{tabular}{llllll}
\hline
\textbf{Measure}                                      & \textbf{Group} & $\mathbf{\overline{x}}$  & \textbf{s} & \textbf{max} & \textbf{min} \\ \hline
Elapsed seconds                   & 3DShEx                          & 417.8                   & 173.034                     & 644                           & 247                           \\
                                  & RDFShape                        & 265.6                   & 106.746                     & 456                           & 204                           \\
                                  & Shumlex                         & 186.5                   & 91.799                      & 314                           & 95                            \\
Success rate                      & 3DShEx                          & 0.583                   & 0.096                       & 0.667                         & 0.5                           \\
                                  & RDFShape                        & 0.7                     & 0.14                        & 0.833                         & 0.5                           \\
                                  & Shumlex                         & 0.708                   & 0.21                        & 1                             & 0.5                           \\
Precision                         & 3DShEx                          & 0.154                   & 0.0757                      & 0.256                         & 0.074                         \\
                                  & RDFShape                        & 0.268                   & 0.074                       & 0.357                         & 0.174                         \\
                                  & Shumlex                         & 0.476                   & 0.365                       & 1                             & 0.151                         \\ \hline
\end{tabular}
\end{table}

Descriptive statistics of the qualitative results are shown in Table \ref{t6} in Appendix A. Overall, user ratings are positive for all tools: 71.1\% of answers express either high or very high level of approval. Tools score on average neutral or positive ratings for every measure, with the exception of Scalability which obtains neutral or negative ratings on average. Statistical analysis showed no significant differences between the three groups for any measure.

\subsection{Discussion}\label{s6.3}
Results for the first test case do not show any significant difference between groups for any of the metrics. This can be explained by the great variability in all groups -e.g. elapsed time for Shumlex ranges from $\sim$1m to $\sim$5m- as well as little difference between means. 

Nonetheless, it should be noted that only one member of the Shumlex group achieved a perfect score both in success rate and precision. Overall, success rate is likely negatively influenced by the scarce theoretical knowledge of Shape Expressions which participants assessed. This would explain how a few simple tasks seem to cause general confusion. Most notably, question 9 which involved pointing out the reference which connected two shapes got no correct answers from both 3DShEx and RDFShape groups, while most Shumlex users answered correctly. Given its uniqueness within the experiment, this particular difference in performance may be due to either an underlying cause or pure chance.

Thus, it may be only stated with certainty that \textbf{in cases with low diagrammatic complexity, there is no evidence of difference in performances between tools}. Since only the Shumlex group managed to complete all the proposed tasks and performed adequately in error-prone tasks, there may be need of further evaluation with larger samples to assess the potential influence of the tool in success rates and precision.

Regarding the second test case, \textbf{elapsed times show significant differences between groups} with a large effect size ($\eta^2$=0.405). Post-hoc results suggest that 3DShEx users require more time to perform tasks on large cases than Shumlex users. This may be explained by the combination of a novel navigable space with large diagrammatic complexity causing high cognitive load, thus exceeding working memory limits. This high cognitive load hypothesis is supported by the fact that the only 3DShEx user with lower spatial ability obtained the highest time in the experiment. As stated in Section \ref{s3.1}, lower spatial ability may imply higher cognitive loads in 3D enviroments. Superior time performance for Shumlex users may be due to its closer resemblance to UML class diagrams, whose specification users claimed to be familiar with.

\begin{figure}[h]
\includegraphics[scale=0.4]{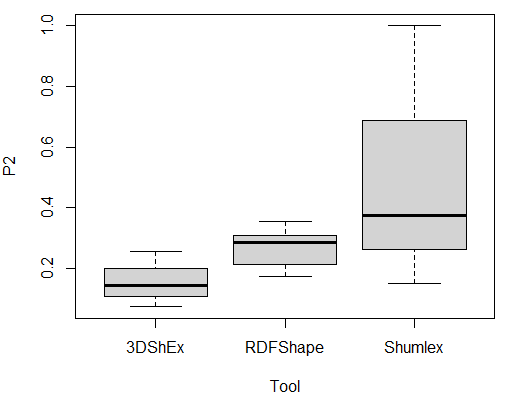}
\caption{Precision values by tool in the second test case.}\label{f6}
\end{figure}

Finally, results show no significant difference between groups at any of the variables in the subjective evaluation performed by the students. Average ratings are mostly positive or neutral if not. The sole exception in the qualitative analysis is the variable Scalability\footnote{"The tool is most useful in large use cases."}, where the tools scored either neutrally or negatively. 
This suggests that users perceive the visualization tools to be of more use with small or medium schemas. Even though complexity management tools seem to be appreciated, extraneous cognitive load may be still excessive. This effect is probably exacerbated by their unfamiliarity with the language and the notation.

Oddly enough, while RDFShape scores a 3.8/5 in Complexity Management, when asked for feedback some of its users convey naught but dissatisfaction in this regard. \emph{"As expected, the larger the use case, the more confusing the diagram"} or \emph{"In very large graphs it is complicated to see the arrows that link entities in the central regions"}. In spite of those -expected- statements, neither rates it negatively. 

In the light of such contradictions, there are several possible explanations. They might have not completely understood the statement to assess\footnote{"The tool facilitates the understanding of complex areas."}, or they might have feared that too harsh an assessment would be detrimental to our interests (a RDFShape user even gives a perfect score to all variables but one). Mayhap it is merely a consequence of their lack of experience. Either way, it may seem like user feedback holds information of greater value to us.

Further analysis of user's comments unveils a similar perception of 3DShEx's complexity management. \emph{"In small cases it is very useful, in large cases like the second one it is quite difficult to deal with"} and \emph{"In the second case it was impossible to follow the relationships and to find the texts of the relationships for each shape"}. It cannot be concluded whether they actually used the complexity management mechanisms. However, if that is the case those may be unintuitive to the users; either way, the tool fails at meeting those needs. 

By comparison, there is a single comment related to Shumlex: \emph{"In the second case, [...] it can be a little complicated to discern the name of the relationship. You can select the shape from which it comes out to differentiate [the name] but it would be nice to be able to do it by clicking on it or hovering over it"}. The contribution of the mechanism is appreciated while providing alternative solutions to that particular task.

As a summary of the qualitative analysis, \textbf{complexity management perception seems to be more favourable to Shumlex}, while the remaining variables appear to have a similar impact throughout the tools.

\section{Conclusions and future work}\label{s7}
A UML-based visual notation for ShEx has been proposed, which is built upon broadly used and operationalized principles. Moreover, said notation has been implemented in both 2D and 3D prototypes, named respectively \emph{Shumlex} and \emph{3DShEx}. Results of both qualitative and quantitative analysis lead to the following conclusions:
\paragraph{Efficiency of Shumlex.} Even though Shumlex users mostly obtained better results than participants with other tools independently of the test case, the small sample size implies that those differences were not significant enough as to be able to generalize claims of efficiency. Nonetheless, the above together with receiving the most positive user feedback make us think that such universalization may be possible with further research. The absence of widespread complaints about complexity management -as it occurs in the others- is likely to be the result of the mechanisms put in place.
\paragraph{Cognitive overload in 3D environment.} 3DShEx users were significantly slower than Shumlex users in a large use case. Furthermore, 3DShEx obtains the worst average ratings in success rate, precision and most qualitative variables. While differences are not statistically significant in those, it is considered  likely that further research may provide a basis for confirmation. Lastly, despite having at their disposal a complexity management mechanism similar to that of Shumlex, user complaints are directed towards scalability. Given that intrinsic cognitive load is the same as Shumlex, it is concluded that interaction with the 3D environment is causing a greater extraneous cognitive load upon the user. The resulting cognitive overload frustrates the user to the detriment of their comprehension and proper use of the available features.
\paragraph{UML-like visual notation.} By shaping the visual notation to resemble UML class diagrams, it was  hoped to achieve a intuitive, transparent solution without forsaking efficiency. User evaluation of learning ease and semantic transparency is reasonably affirmative of such intent. On the other hand, its efficiency seems rather dependent on the manner in which the visual notation is presented.
\paragraph{Future work.} It is considered that future efforts should be focused on Shumlex as the more promising approach. Analysis of user feedback suggests an extension of the complexity management capabilities as to support more specialized tasks. E.g., being able to select a single shape reference. Moreover, the inclusion of a search engine could be of assistance to users when navigating large schemas.

\begin{appendix}

\section{Qualitative analysis results}

\begin{table}[H]
\caption{Descriptive statistics for qualitative analysis results.} \label{t6}
\begin{tabular}{llllll}
\hline
\textbf{Measure}                                            & \textbf{Group} & $\mathbf{\overline{x}}$  & \textbf{s} & \textbf{max} & \textbf{min} \\ \hline
General satisfaction              & 3DShEx                          & 3                       & 0.82                        & 4                             & 2                             \\
                                  & RDFShape                        & 4.4                     & 0.55                        & 5                             & 4                             \\
                                  & Shumlex                         & 4.25                    & 0.96                        & 5                             & 3                             \\
Ease of use                       & 3DShEx                          & 3.75                    & 1.5                         & 5                             & 2                             \\
                                  & RDFShape                        & 4.2                     & 0.45                        & 5                             & 4                             \\
                                  & Shumlex                         & 4.5                     & 1                           & 5                             & 3                             \\
Learnability                      & 3DShEx                          & 3.75                    & 0.5                         & 4                             & 3                             \\
                                  & RDFShape                        & 4                       & 1.22                        & 5                             & 2                             \\
                                  & Shumlex                         & 4.5                     & 0.58                        & 5                             & 4                             \\
Semantic Transparency             & 3DShEx                          & 3.75                    & 0.5                         & 4                             & 3                             \\
                                  & RDFShape                        & 3.8                     & 0.45                        & 4                             & 3                             \\
                                  & Shumlex                         & 4.25                    & 0.5                         & 5                             & 4                             \\
Applicability                     & 3DShEx                          & 3.25                    & 0.96                        & 4                             & 2                             \\
                                  & RDFShape                        & 4.2                     & 0.45                        & 5                             & 4                             \\
                                  & Shumlex                         & 4                       & 1.15                        & 5                             & 3                             \\
Error proneness                   & 3DShEx                          & 3.25                    & 0.96                        & 4                             & 2                             \\
                                  & RDFShape                        & 3.6                     & 1.14                        & 5                             & 2                             \\
                                  & Shumlex                         & 3.5                     & 1.73                        & 5                             & 2                             \\
Complexity Management             & 3DShEx                          & 3.5                     & 1.29                        & 5                             & 2                             \\
                                  & RDFShape                        & 3.8                     & 0.84                        & 5                             & 3                             \\
                                  & Shumlex                         & 3.25                    & 0.96                        & 4                             & 2                             \\
Scalability                       & 3DShEx                          & 1.25                    & 0.5                         & 2                             & 1                             \\
                                  & RDFShape                        & 3                       & 1.22                        & 5                             & 2                             \\
                                  & Shumlex                         & 2.5                     & 1.29                        & 4                             & 1                             \\
References                        & 3DShEx                          & 3.75                    & 0.96                        & 5                             & 3                             \\
                                  & RDFShape                        & 4                       & 0.71                        & 5                             & 3                             \\
                                  & Shumlex                         & 3.75                    & 0.96                        & 5                             & 3                             \\
Constraints                       & 3DShEx                          & 3                       & 0                           & 3                             & 3                             \\
                                  & RDFShape                        & 3.6                     & 1.14                        & 5                             & 2                             \\
                                  & Shumlex                         & 3.75                    & 0.96                        & 5                             & 3                             \\
Global                            & 3DShEx                          & 3.26                    & 0.8                         & 5                             & 1                             \\
                                  & RDFShape                        & 3.86                    & 0.82                        & 5                             & 2                             \\
                                  & Shumlex                         & 3.83                    & 1.01                        & 5                             & 1                             \\ \hline
\end{tabular}
\end{table}

\section{Metric of similarity}\label{metricannex}
Given a notation \emph{N} with $G_N$ graphical symbols, the operationalization framework of Störrle et al. \cite{oppar} proposes 4 criteria to ascertain perceptual discriminability: visual distance (VD), redundant coding (RC), perceptual pop-out (PPO) and textual differentiation (TD). These are normalized to an interval of [0, 1], in such a way that 0 denotes null discriminability and 1 compliance with all criteria. Finally, the average discriminability for the notation is calculated.
\subsection{Visual distance}
Both the visual variable difference function \emph{vvd(a,b)} and the visual distance function \emph{vd(g,h)} are used as-is. Maximum value of \emph{vd(g,h)} is 1 and there are $|G_N^2|$ posible combinations of \emph{g,h}. Hence, the average visual distance function \emph{VD(N)} uses $|G_N^2|$ as a denominator so that 1 is the highest value possible. This does not take into account that whenever \emph{g} equals \emph{h, vd(g,h) = 0}, so the maximum value of the summation is $|G_N^2| - |G_N|$. Thus, the denominator is modified and the substraction of the unit removed so that the values are normalized in the specified range.
\[VD(N) := \frac{1}{|G_N^2| - |G_N|} \sum_{g,h \in G_N} vd(g,h)\]
For instance, the visual distance between the graphical symbols \emph{box (b)} and \emph{directed arrow (da)} is as follows. As suggested, weights \emph{w} are 7 for shape and 1 for the rest of visual variables. Those visual variables not used have \emph{vvd = 0}, thus only shape, brightness and texture may be computed. In this particular case, shapes are in different main groups: lines and regions. Therefore, $vvd(v_{sh}(b),v_{sh}(da)) = 1$. Same for brightness, given that the colour of their main areas is completely opposite (black and white). However, they do have the same solid texture, hence $vvd(v_{tx}(b),v_{tx}(da)) = 0$.
\[vd(b,da) := \frac{1}{||w||} \sum_{i=1}^{d} w_{i} \cdot vvd(v_{i}(b),v_{i}(da)) := \frac{1}{14} (7\cdot1+1\cdot1+1\cdot0) := 0.57\]
This same procedure is repeated for all combinations of graphical symbols. The only new development is the comparison between shapes of the same basic group (i.e. arrows and lines) for which \emph{vvd = 0.5}. Its results are shown in Table \ref{t4}.

\begin{table}[H]
\caption{Visual distance for all graphical symbol pairs.} \label{t4}
\begin{tabular}{lllll}
\hline
               & Box  & Directed arrow & Diamond arrow & Dashed line \\ \hline
Box            & 0    & 0.57           & 0.57          & 0.64        \\
Directed arrow & 0.57 & 0              & 0.32          & 0.32        \\
Diamond arrow  & 0.57 & 0.32           & 0             & 0.39        \\
Dashed line    & 0.64 & 0.32           & 0.39          & 0           \\ \hline
\end{tabular}
\end{table}

With all of the above, VD for our notation N may be finally calculated. 
\[VD(N) := \frac{1}{16 - 4} (0.57\cdot4+0.32\cdot4+0.64\cdot2+0.39\cdot2) := 0.47\]
\subsection{Redundant coding}
Previous changes to the denominator apply to the RC(N) function as well. The \emph{vr(g,h)} function is used as-is. Results are displayed in Table \ref{t5}.

\begin{table}[H]
\caption{Redundant coding for all graphical symbol pairs.} \label{t5}
\begin{tabular}{lllll}
\hline
               & Box  & Directed arrow & Diamond arrow & Dashed line \\ \hline
Box            & 0    & 0.25           & 0.25          & 0.38        \\
Directed arrow & 0.25 & 0              & 0.25          & 0.25        \\
Diamond arrow  & 0.25 & 0.25           & 0             & 0.38        \\
Dashed line    & 0.38 & 0.25           & 0.38          & 0           \\ \hline
\end{tabular}
\end{table}
\[RC(N) := \frac{1}{16 - 4} (0.25\cdot8+0.38\cdot4) := 0.29\]
\subsection{Perceptual popout}
Each graphical symbol has at least a unique value in one visual variable. Taking into account the \emph{shape} variable alone fulfils this criterion. Once more, the substraction is removed so that the best possible value is 1. Therefore, the function is as follows:
\[PPO(N) := \frac{1}{4} (1\cdot4) := 1\]
\subsection{Textual differentiation}
As stated in Section \ref{semioticclarity}, PoN does not consider visual constructs that make use of textual differentiation to convey distinct meanings as different graphical symbols (symbol overload). Therefore, a modification is made to this criterion so that it does not measure the proportion of graphical symbols which only differ by textual cues. The proportion of graphical symbols which convey several concepts by textual differentiation is evaluated instead.
\[TD(N) := 1 - \frac{2}{4} := 0.5\]

\end{appendix}

\nocite{*}
\bibliographystyle{ios1}           
\bibliography{bibliography}        

\end{document}